\documentclass[%
 reprint,
superscriptaddress,
 amsmath,amssymb,
 aps,
]{revtex4-1}
\usepackage{graphicx}
\usepackage{amsmath}
\usepackage{float}
\usepackage{comment}
\usepackage{amsmath}
\usepackage{amssymb}
\usepackage{mathrsfs}
\usepackage[below]{placeins}
\usepackage{fancyhdr}
\usepackage{soul,color}

\begin{document}


\title[\texttt{achemso} demonstration]
{Charge Order at High Temperature in Cuprate Superconductors}


\author{Riccardo Arpaia}
\email{riccardo.arpaia@chalmers.se}
\affiliation{Quantum Device Physics Laboratory, Department of Microtechnology and Nanoscience, Chalmers University of Technology, SE-41296 G\"{o}teborg, Sweden}
\author{Giacomo Ghiringhelli}
\email{giacomo.ghiringhelli@polimi.it}
\affiliation{Dipartimento di Fisica, Politecnico di Milano, Piazza Leonardo da Vinci 32, I-20133 Milano, Italy}
\affiliation{CNR-SPIN, Dipartimento di Fisica, Politecnico di Milano, Piazza Leonardo da Vinci 32, I-20133 Milano}

\date{\today}
\begin{abstract}
The presence of different electronic orders other than superconductivity populating the phase diagram of cuprates suggests that they might be the key to disclose the mysteries of this class of materials. In particular charge order in the form of charge density waves (CDW), i.e., the incommensurate modulation of electron density in the CuO$_2$ planes,  is ubiquitous across different families and presents a clear interplay with superconductivity. Until recently, CDW had been found to be confined inside a rather small region of the phase diagram, below the pseudogap temperature and the optimal doping. This occurrence might shed doubts on the possibility that such ``low temperature phenomenon'' actually rules the properties of cuprates either in the normal or in the superconducting states. However, recent resonant X-ray scattering (RXS) experiments are overturning this paradigm. It results that very short-ranged charge modulations permeate a much wider region of the phase diagram, coexisting with CDW at lower temperatures and persisting up to temperatures well above the pseudogap opening. Here we review the characteristics of these high temperature charge modulations, which are present in several cuprate families, with similarities and differences. A particular emphasis is put on their dynamical character and on their coupling to lattice and magnetic excitations, properties that can be determined with high resolution resonant inelastic x-ray scattering (RIXS).
\end{abstract}

\pacs{}

\maketitle

\section{Introduction}
\begin{figure}[b]
\centering
\includegraphics[width=5.2cm]{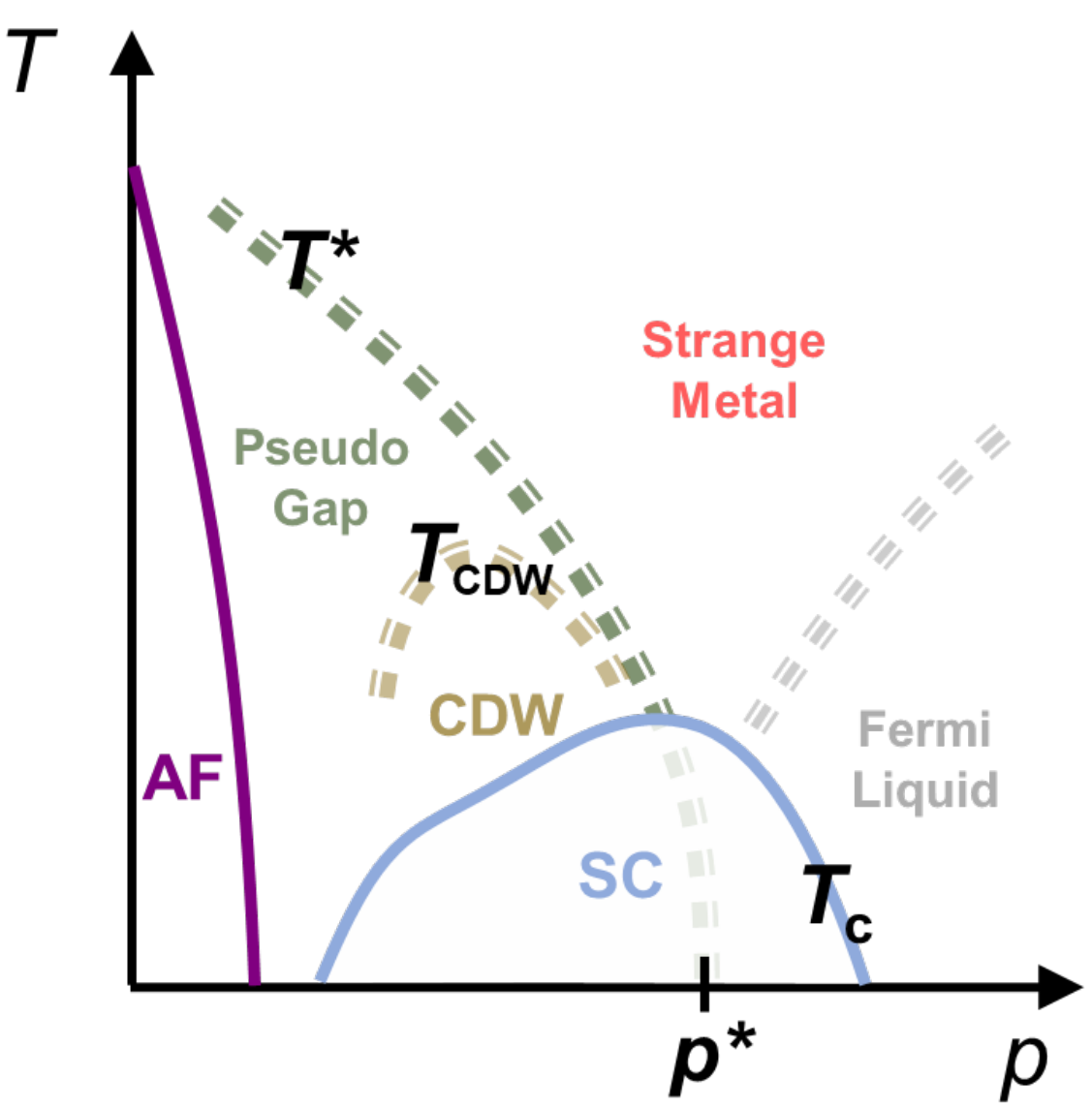}
\caption{{\footnotesize Temperature vs hole doping phase diagram of cuprates, as it looked before the recent introduction of high temperature charge density fluctuations. Here charge order, in the form of charge density waves (CDW), manifests only in a confined region of the pseudogap regime, below $T_{\mathrm{CDW}}$. AF and SC indicate respectively the antiferromagnetic and the superconducting regions. $T^{\mathrm{*}}$ and $T_{\mathrm{c}}$ are respectively the pseudogap and the superconducting critical temperature. $p^{\mathrm{*}}$ is the putative quantum critical point, at $p \approx 0.19$.}} \label{fig:Fig1}
\end{figure}

The temperature-doping ($T$-$p$) phase diagram of the high critical temperature cuprate superconductors (HTS) (see Fig. \ref{fig:Fig1}) is characterized by a plethora of ordered states, whose properties are tuned by doping and driven by many competing degrees of freedom (charge, spin, lattice and orbitals) \cite{keimer2015quantum, fradkin2015colloquium}. The comprehension of the intertwining/competition between these ordered states is a crucial step toward the understanding of two of the grand challenges in solid state physics: high-temperature superconductivity, and the strange metal phase. The main puzzle is that HTS seem to obey the Bardeen-Cooper-Schrieffer (BCS) quasiparticle theory, when they are superconducting, but the normal state is undoubtedly different from that of a Fermi Liquid on which BCS theory is based. Indeed, the resistance shows an anomalously linear behavior vs temperature that, for optimally doped samples ($p \approx 0.17$, with $p$ hole doping for planar copper atom), dominates down to the superconducting critical temperature $T_{\mathrm{c}}$ \cite{ando2004electronic, barivsic2013universal}. The way the various underlying quantum electronic orders intertwine and are eventually responsible for the phenomenology in HTS compounds is unknown. Many theoretical proposals have been suggested so far, but none of them composed the puzzle in its entirety. New experiments, facing these complexities from a different perspective, are needed in order to clarify the physics at play. 

Among the ordered states, mainly developing below a doping-dependent pseudogap temperature, where states are lost at the Fermi energy, a special role is played by charge order. This electronic phase is characterized by the self-organization of the valence electrons in the CuO$_2$ planes into periodic structures, incompatible with the periodicity of the underlying lattice and therefore breaking the translational symmetry. Charge modulations were first detected indirectly by neutron scattering in 1995 in the form of stripes, where charge and spin modulations are coupled \cite{tranquada1995evidence}. For almost 15 year after such discovery, limited to the La-based cuprate family, the existence in other families of charge modulations has been elusive and/or observed{, more indirectly, via Scanning Tunneling Microscopy (STM) experiments \cite{hoffman2002four, vershinin2004local, hanaguri2004checkerboard, wise2008charge, fujita2011spectroscopic}}. In 2011 a resonant X-ray scattering (RXS) work demonstrated the presence of incommensurate charge order in YBa$_2$Cu$_3$O$_{7-{\delta}}$ (YBCO), in the form of charge density waves (CDW), independent of any spin modulations \cite{ghiringhelli2012long}. These CDW, which are quasi-2D (i.e. the correlation length along  $c$-axis is negligible), are located in the underdoped region of the phase diagram (e.g., $0.08 < p < 0.17$ for YBCO), below  an onset temperature $T_\mathrm{CDW}$, which is lower than the pseudogap temperature $T^*$ (see Fig. \ref{fig:Fig1}).  In the same months indirect observation of CDW in YBCO came also from nuclear magnetic resonance \cite{wu2011magnetic} and high energy X-ray diffraction \cite{chang2012direct}. After the first discovery, other RXS experiments showed that CDW are present, with minor differences, in all major hole doped { and electron doped} cuprate compounds: YBCO \cite{chang2012direct, achkar2012distinct, blanco2013momentum, blanco2014resonant}, Bi$_2$Sr$_{2-x}$La$_{x}$CuO$_{6+\delta}$ (Bi2201) \cite{he2014fermi, comin2014charge, peng2016direct}, Bi$_2$Sr$_2$CaCu$_2$O$_{8+\delta}$ (Bi2212) \cite{da2014ubiquitous, chaix2017dispersive}, HgBa$_2$CuO$_{4+\delta}$ (Hg1201) \cite{tabis2014charge, tabis2017synchrotron}, { Nd$_{2-x}$Ce$_{x}$CuO$_{4}$ (NCCO) \cite{da2015charge, jang2017superconductivity, da2018coupling}}. It was quickly demonstrated that charge density waves are ubiquitous in HTS.

Where does the interest for CDW in HTS raise from? Indeed, the concept of charge modulation is not new: evidences for an order of such kind, associated to a reconstruction of the Fermi surface, have been found since the 1950's in many binary and ternary compounds with weakly correlated quasi-1D or quasi-2D electronic structure \cite{wilson1975charge, gibbs1988high, monceau2012electronic}. Nevertheless, in all these compounds the charge order instability can be taken into account within the conventional picture for superconducting and normal metal state. In HTS, instead, the role played by charge order might be more profound and possibly linked or even responsible for the unconventional normal state and superconducting properties. In cuprates the charge order exhibits indeed a strong entanglement with the superconducting order, highlighted by many occurrences, as the depression of the superconducting critical temperature at the doping where charge order is strongest ($p \approx 1/8$) \cite{ghiringhelli2012long, blanco2014resonant, comin2016resonant}, the drop of the CDW intensity and correlation length when entering in the superconducting state \cite{ghiringhelli2012long, chang2012direct, huecker2014competing, blanco2014resonant, da2014ubiquitous}, the rise of these two quantities - together with the appearance of a long range three-dimensional CDW (3D-CDW) - when superconductivity is depressed under high magnetic fields \cite{wu2011magnetic, chang2012direct, gerber2015three}. In particular the 3D-CDW, characterized by a correlation length diverging at very low temperatures, would point toward to existence of a quantum critical point driven by charge order at a doping level close to the optimally one. 
Thence, it is not surprising that in several theoretical models, initially proposed by Emery {\itshape et al} \cite{emery1990phase} and Castellani {\itshape et al} \cite{castellani1995singular} and variously developed in the ensuing two decades \cite{emery1993frustrated, low1994study, castellani1996non, perali1996d, kivelson2003detect, kloss2016charge, montiel2017effective, caprara2017dynamical, caprara2017pseudogap, caprara2019ancient, di2020revival}, collective charge  density  fluctuations are expected to permeate a very broad area of the cuprate phase diagram, being pivotal both to the  anomalous properties of the normal state and to the superconducting pairing. 

However, despite the intriguing experimental observations and theoretical expectations, the ability of charge order to influence the properties of HTS has been questioned for a long time. This is because CDW are localized  in a rather narrow region of the phase diagram below the pseudogap temperature: they are absent in strongly underdoped and overdoped samples, even though superconductivity sets in at these doping levels; they are absent at high temperatures, where the strange metal phase dominates the phase diagram. So, if keeping into account only CDW, it could be easier to consider charge order as a mere epiphenomenon on top of a fundamentally peculiar metallic state rather than as a main player in the HTS phenomenology \cite{anderson1990luttinger, wen1996theory, rice2011phenomenological, sachdev2016spin, cai2017intertwined}.

This scenario has been overturned thanks to recent resonant X-ray scattering experiments. By taking advantage of the superior sensitivity of modern Resonant Inelastic X-ray Scattering (RIXS) facilities, several studies have been done on different cuprate families, by thoroughly investigating  the temperature and doping dependence of charge order. The result is that {\itshape charge density modulations, with a shorter correlation length with respect to low temperature CDW, are present at any investigated temperature and in a broad doping range} \cite{arpaia2019dynamical, yu2020unusual, wang2020doping, lee2021spectroscopic, boschini2021dynamic, miao2017high, miao2019formation, wen2019observation, miao2021charge, lin2020strongly, wang2020high}. Consequently, they persist  at temperatures exceeding not only the previously-defined $T_\mathrm{CDW}$, but also $T^*$. This occurrence is universal for cuprates, since it has been observed in all the families of investigated compounds, including those with the low temperature locking of spin and charge modulations. In view of these experimental results, the $T$-$p$ phase diagram gets profoundly reshaped: in order to explain the different orders and regions characterizing it, one cannot ignore the presence of these pervasive high temperature charge fluctuations. Several proposals, linking these fluctuations either to the superconducting pairing mechanism \cite{yu2020unusual} or to the strange metal phase \cite{seibold2021strange, caprara2020dissipation}, have already been formulated.

Even though there is still a lot to understand about this novel phenomenology, in this review we want to summarize all the recent RXS results, pointing to the presence of high temperature charge modulations and to their universality in cuprate HTS. We will focus on the characteristics of these modulations: their short correlation length and their dynamics. In particular, the finite energies characterizing them highlight a strong intertwining between charge order and other excitations. We will therefore spotlight the anomalies occurring in the dispersion of various optical phonons and in the high-energy magnetic excitation spectrum, which appear related to charge modulations. Finally, we will face the issue of the universality of the phenomenon, showing that, although with some differences, charge order is present at high temperature also in the systems where the CDW are strongly modified or missing: the La-based ``striped" family and the electron-doped cuprates.  

Last but not least, we want to point out that the understanding of the charge order phenomenon in cuprates and of its characteristics is rapidly evolving and significantly expanding. This is driven by the recent technological advancement in X-ray scattering techniques, and by the increased accessibility now available worldwide to new-generation RIXS facilities. Such fast progress calls for this review, which covers the most significant aspects on charge modulations which have been discovered in the last two, three years. For brevity, the focus will be  here on X-ray scattering experiments, which have brought in most of the recent results. However, we note that  in the last decade also other techniques, such as ultrafast spectroscopy \cite{torchinsky2013fluctuating} and neutron scattering \cite{park2014evidence, jacobsen2018distinct}, have provided evidence of high-temperature, dynamical charge order in cuprates. For a complete and detailed overview of CDW and 3D-CDW results we refer to the articles by Comin \cite{comin2016resonant} and Frano  \cite{frano2020charge}.


\section{At the origin of the ``charge order as low-temperature phenomenon'' paradigm}
\begin{figure*}[t]
\centering
\includegraphics[width=16cm]{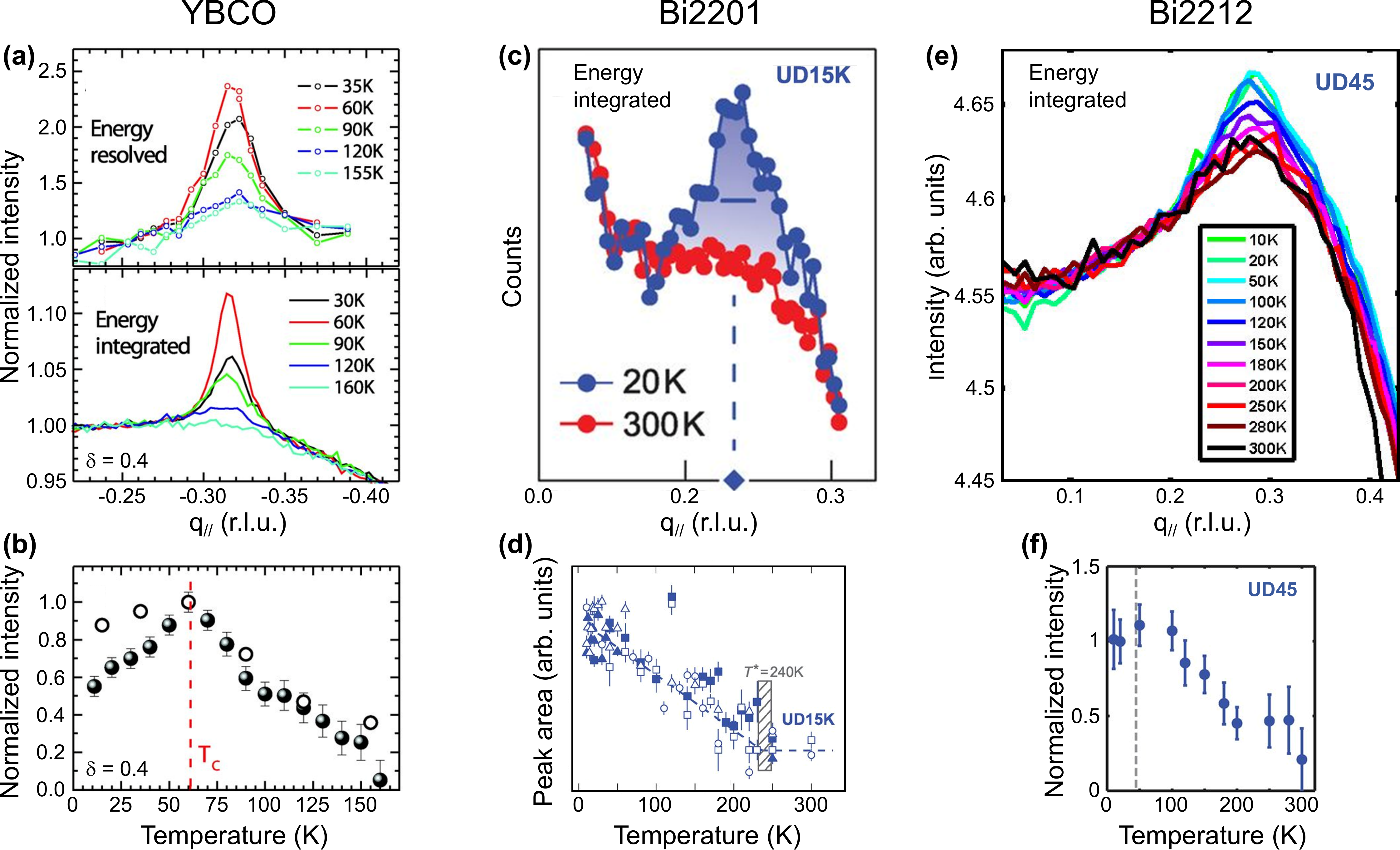}
\caption{{\footnotesize Temperature dependence of the CDW signal measured on several underdoped cuprates in pioneering resonant X-ray scattering works. (a) $q$ scans measured as a function of the temperature on a YBa$_2$Cu$_3$O$_{7-{\delta}}$ (YBCO) single crystal using both energy resolved resonant X-ray scattering (RIXS) and energy integrated resonant X-ray scattering (EI-RXS). {Adapted from Ref. \citenum{ghiringhelli2012long} (\copyright 2012 
American Association for the Advancement of Science)}. (b) $T$-dependence of the CDW intensity derived from the energy-resolved (open circles) and the energy-integrated data (solid circles) of panel (a). { Adapted from Ref. \citenum{ghiringhelli2012long} (\copyright 2012 
American Association for the Advancement of Science)}. (c)-(d) Same as (a)-(b), but the energy integrated data have been acquired on a Bi$_2$Sr$_{2-x}$La$_x$CuO$_{6+\delta}$ sample. {Adapted from Ref. \citenum{comin2014charge} (\copyright 2014 
American Association for the Advancement of Science)}. (e)-(f) Same as (a)-(b) and (c)-(d), with measurements here performed on Bi$_2$Sr$_2$CaCu$_2$O$_{8+x}$. {Adapted from Ref. \citenum{da2014ubiquitous} (\copyright 2014 
American Association for the Advancement of Science)}. To reveal the presence  of  the  CDW  order  component and determine its temperature dependence, the  RIXS and/or EI-RXS  scans  at  different  temperatures  are subtracted by a reference scan measured at high temperature, since in that temperature range the scans are almost unchanged. As a consequence, although peaked at an in-plane wavevector $q_{//}$ similar to that of CDW, this scan is used as a background. This procedure intrinsically hides the presence of any charge orders at high temperature.}} \label{fig:Fig2}
\end{figure*}

Already in the first, pioneering, RXS experiment showing the presence of CDW in a cuprate system, the temperature dependence of the CDW peak  was thoroughly investigated (see Figs. \ref{fig:Fig2}(a)-(b)) \cite{ghiringhelli2012long}. 
There, the CDW onset temperature of a YBa$_2$Cu$_3$O$_{6.6}$ single crystal, i.e. at a doping level where the CDW is strongest and $T_{\mathrm{CDW}}$ is maximum, is estimated to occur in the range $\sim$150-160 K. This temperature is significantly lower than the pseudogap temperature at that doping level ($T^* \sim$ 210\,K, as one can indirectly extract from transport experiments \cite{barivsic2013universal, arpaia2018probing}). Such low $T_{\mathrm{CDW}}$ is determined from both resonant inelastic X-ray scattering (RIXS) and energy-integrated RXS (EI-RXS) measurements. Above this temperature, the $q$-scans measured along the ($H,0$) direction become nearly $T$-independent, still exhibiting however a peaked shape, with the maximum intensity in correspondence of the CDW wave vector $q_{\mathrm{CDW}}$ (see curves in Fig. \ref{fig:Fig2}(a) {related to 155 K and 160 K}). This is a hint of a possible high-temperature charge density modulation signal, but its presence is not discussed in the manuscript. The reason for that is mainly related to the subtraction procedure used to single out the CDW peak from each $q$ scan, which is a direct consequence of the rather low probing sensitivity of the EI-RXS scans. In these scans the  elastic/quasi  elastic  component, whose changes are directly connected with the presence of charge order, is indeed affected by a large  background, added by the detector and connected  to  the  inelastic  component. If present, a very small variation of the elastic component at temperatures above $T_{\mathrm{CDW}}$ could be therefore hidden by the large inelastic background. For the same reason, any $q$-dependence of the inelastic scattering might be misinterpreted as an elastic signal due to charge order. Therefore the safest procedure, developed to reveal the presence  of  the  CDW  component, is the following: {\itshape the  EI-RXS scans  at  different  temperatures  are subtracted by a reference scan measured at the threshold (high) temperature, above which the scans appear unchanged. This procedure intrinsically cancels out  any charge order signal possibly present at and above the threshold temperature}. 
In RIXS, the probing sensitivity is higher, since by measuring with a spectrometer the full spectra at each momentum point one can easily isolate the elastic/quasi elastic signal coming from charge order from the dominant inelastic contribution. This superior selectivity comes at the price of a lower signal and more complicated instrumentation. RIXS experiment are indeed slower than EI-RXS and suffer from a lower statistical quality, although eventually the higher selectivity carries decisive advantages when it comes to the interpretation of the results. For both techniques, RIXS and EI-RXS, measurements benefit from diffraction-quality sample manipulation and cryo-control down to the lowest possible temperatures. It must be noted that RIXS energy resolution has to be pushed as much as possible to get the best insight from the data. In the first RIXS experiment on CDW the energy resolution was $\sim 130$ meV, enough to cut out the strong $dd$ and charge transfer spectral contributions, but insufficient to fully subtract the spin and phonon peaks. This explains why in Reference  \citenum{ghiringhelli2012long}, the higher sensitivity of RIXS was not fully trusted when it detected  possibly persistent charge order signal at high temperatures not seen by EI-RXS  (see empty circles in Figure \ref{fig:Fig2}(b)). Nevertheless, it was already evident that RIXS measurements were richer in information and more sensitive than EI-RXS.

Since EI-RXS became the main technique to study the temperature evolution of the CDW peak, because of its fast acquisition times and broader availability of instrumentation and beam time, such subtraction procedure started to be routinely used in other works, focusing both on YBCO and on other cuprate families. Ref.~\citenum{blanco2014resonant} presents a very detailed EI-RXS investigation of the CDW in YBCO as a function of the doping, showing for the first time the complete dome of $T_{\mathrm{CDW}}$: it lays entirely in the pseudogap region, at doping levels $0.08 < p < 0.16$, with the onset temperature ranging from $\sim 110$ to 160\,K depending on $p$. Above this temperature, a very broad, temperature-independent peak is still visible, centered at approximately the same wavevector $q_{\mathrm{CDW}}$, but it is withdrawn by the subtraction procedure.

A similar broad-in-$q$ peak survives at very high temperatures also in the Bi-based compounds. In Bi$_2$Sr$_{2-x}$La$_x$CuO$_{6+\delta}$ (see Figs. \ref{fig:Fig2}(c)-(d)) \cite{comin2014charge}, the EI-RXS scans measured along the crystallographic axes change with temperature in a range going from 20\,K up to $T_{\mathrm{CDW}}$, which is very close to the temperature corresponding to the opening of the pseudogap, as directly probed by angular resolved photoemission spectroscopy (ARPES). Above $T_{\mathrm{CDW}}$, a broad peak centered at $q_{\mathrm{CDW}}$ is still visible (see Fig. \ref{fig:Fig2}(c)), but it is treated as a featureless background dominated by fluorescence. The same phenomenology applies to underdoped Bi$_2$Sr$_2$CaCu$_2$O$_{8+x}$ samples (see Figs. \ref{fig:Fig2}(e)-(f)) \cite{da2014ubiquitous}. The raw EI-RXS scans are still significantly peaked at 300\,K and $q_{//} \approx q_{\mathrm{CDW}} = 0.29$ r.l.u., even though the signal is almost $T$-independent. To isolate the CDW peak from the raw data without excluding \textit{a priori} the possibility of a high temperature signal, a fourth order polinomial fit is performed on the tails ($q_{//} < 0.22$ r.l.u. and $q_{//} > 0.37$ r.l.u.) of the $q$ scans at different temperatures.  After  subtraction  from  the  fitted  background, each data is fit to a Lorentzian function  in order to  extract  the  temperature dependence of the peak. The result is that a charge order signal cannot be excluded even at 300\,K, although its intensity is rather low, and despite the large errors intrinsic of this subtraction procedure   (see Fig. \ref{fig:Fig2}(f)).

The procedure which takes advantage of the high-temperature measurement as a reference to isolate the low-temperature CDW contribution from the actual background has been used in all the RXS studies performed in the years following the first discovery of CDW in cuprates. This includes investigations in Hg-based \cite{tabis2014charge} and in La-based \cite{fink2009charge} compounds. This has brought to confine CDW in a narrow region of the phase diagram, smaller that it actually covers. And it has been enough to generate the paradigm of the charge order as a ``low-temperature phenomenon''.

This picture has been challenged by more recent experiments, which took advantage of the new generation RIXS facilities that offer better sensitivity, higher energy resolution and diffraction-quality sample handling. This has given indeed the possibility to measure, in a relatively short amount of time, spectra with a statistical quality and resolution that unveiled charge order signals previously lost in the noise.

\section{Dynamical charge density fluctuations} \label{sec:sec3}

\begin{figure*}[htb]
\centering
\includegraphics[width=17cm]{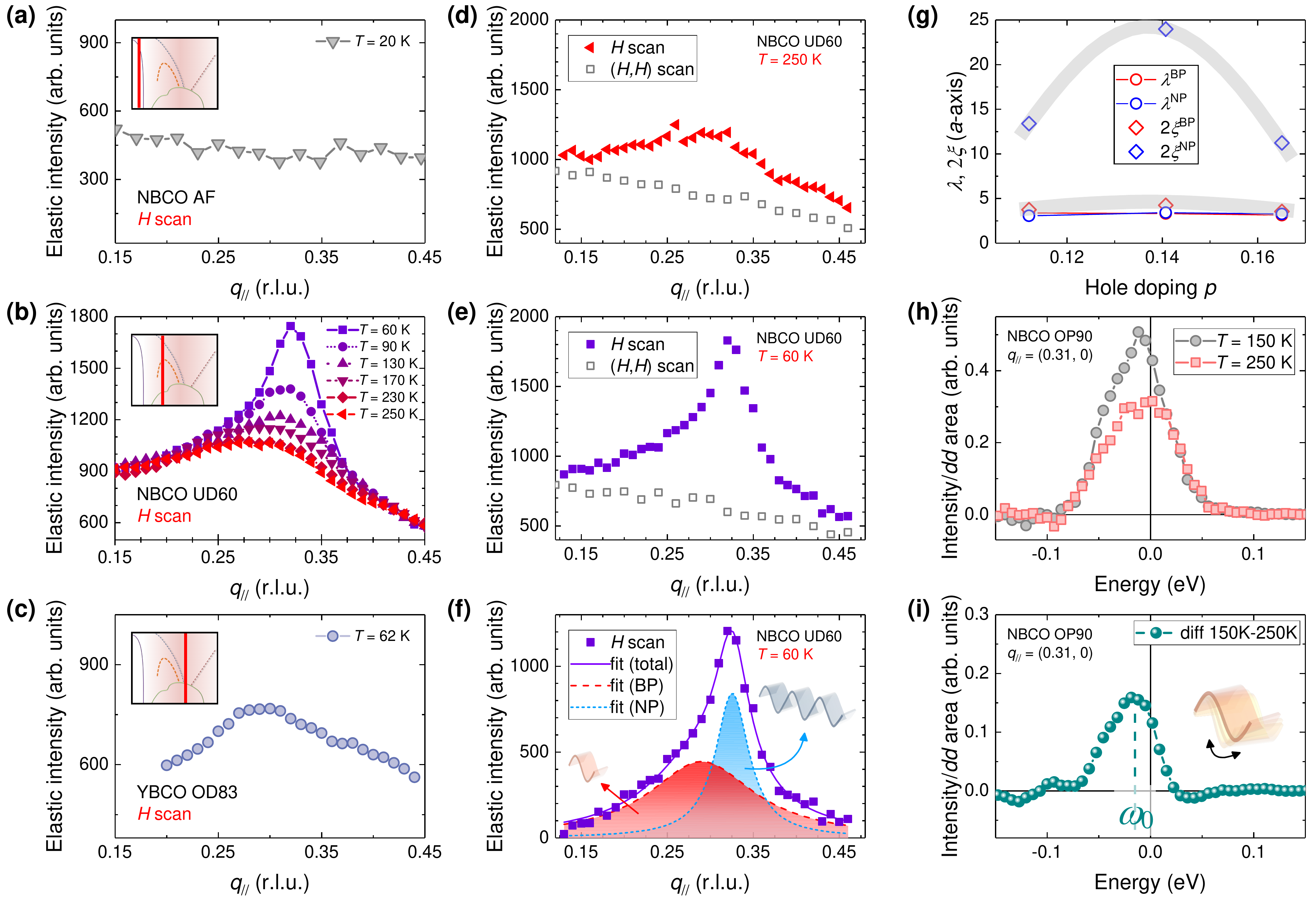}
\caption{{\footnotesize Detection of dynamical charge modulations in (Y,Nd)Ba$_2$Cu$_3$O$_{7-\delta}$ up to room temperature using RIXS. The quasi-elastic intensity of the spectra is investigated as a function of the in-plane wavevector $q_{//}$ and of the temperature in samples with doping going from (a) the antiferromagnetic level ($p \approx$ 0), passing through (b) the underdoped ($p \approx$ 0.11), up to (c) the slightly overdoped regime ($p \approx$ 0.18). While no signal is detected in the antiferromagnetic sample, a broad-in-$q$ peak is clearly present in the underdoped sample up to the highest investigated temperature. A similar broad peak is present in the slightly overdoped sample, where CDW are missing. The broad-in-$q$ peak disappears when moving from the ($H$,0) direction to the Brillouin zone diagonal. The ($H$,$H$) scan can therefore be used as a background both (d) at high temperature and (e) at low temperature. (f) In the latter case, after subtracting the ($H$,$H$) background, the residual peak can be fitted as the sum of two lorentzian: one, narrower, with the typical characteristic of the CDW; the second one, broader, with is associated to dynamical charge density fluctuations. (g) They are fluctuations, since their correlation length $\xi = (\pi \cdot \mathrm{FWHM}_{T_\mathrm{c}})^{-1}$ is of the same order as their wavelength $\lambda = (q_\mathrm{CDW})^{-1}$, differently than CDW, having $\xi >> \lambda$. (h) They are dynamical, since the quasi-elastic region of high-energy resolution RIXS spectra, measured at temperatures where CDF are dominating, are strongly inelastic, even after removing the main phonon contribution.  (i) From the difference between two high-temperature spectra, the energy  of CDF has been determined at the optimally doped level as $\omega_0 \approx 10$ meV. { Adapted from Ref. \citenum{arpaia2019dynamical} (\copyright 2019 
American Association for the Advancement of Science)}.}} \label{fig:Fig3}
\end{figure*}

To properly assess the role of charge density modulations and their extension in the cuprate phase diagram, in Ref. \citenum{arpaia2019dynamical} RIXS measurements on (Y,Nd)Ba$_2$Cu$_3$O$_{7-\delta}$ have been performed over a broad in-plane wave vector $q_{//}$ range as a function of the doping and of the temperature. The samples span from the antiferromagnetic up to the slightly overdoped region \cite{salluzzo2005thickness, baghdadi2014toward, arpaia2018probing, arpaia2019untwinned}, while the temperature from 20\,K up to room temperature. The experiment aimed at retracing the pioneering RXS experiment on YBCO while taking advantage of the superior sensitivity and energy resolution available at the ID32 RIXS beamline at ESRF using the ERIXS spectrometer \cite{brookes2018beamline}.
The results are summarized in Figure \ref{fig:Fig3}.

The quasi-elastic intensity of the spectra, shown in Figure \ref{fig:Fig3}(b) for the underdoped sample ($p = 0.11$), presents a trend versus temperature which strongly resembles that previously observed in Ref. \citenum{ghiringhelli2012long} (see Fig. \ref{fig:Fig2}(a)). A narrow, CDW peak is present, centered at $q_{\mathrm{CDW}} \approx 0.33$ r.l.u. and presenting a strong temperature dependence. At the highest investigated temperatures, an almost $T$-independent broad peak still persists, centered at a slightly different $q_{//}$. A first, strong hint this cannot be a mere background comes from the investigation of the antiferromagnetic sample, presenting  a linear behavior versus $q_{//}$ at all temperatures (see Fig. \ref{fig:Fig3}(a)). The second hint comes instead from the analysis of the  RIXS spectra measured along the Brillouin zone diagonal. There, the quasi-elastic scans show a featureless linear shape at any temperature under investigation (see empty squares in Figs. \ref{fig:Fig3}(d)-(e)). These two occurrences indicate that the quasi-$T$-independent scattering signal present at high temperatures along the ($H$,0) direction is genuine and representative of a short-range charge order. In addition to that, these new findings set out the path for the analysis needed to collect the whole of the charge order scattering signal: the quasi-elastic scan measured along the diagonal, i.e. in the ($H$,$H$) direction, can be used as indicative linear background in the fitting  of  the  scans  along  ($H$,0). After subtracting the ($H$,$H$) scan from the ($H$,0), the resulting curve can be fitted by assuming at high temperatures a single, broad Lorentzian profile. Decreasing the temperatures, the fit can instead  be performed by taking into consideration two Lorentzian profiles centered at similar, but non identical, $q_{//}$ values: one broader, with intensity $I$ and width close to those measured at high temperature, and one narrower, which retains all the characteristics of the previously studied CDW (see Fig. \ref{fig:Fig3}(f)). Notably, the volume, i.e. the total  integrated in-plane scattering  intensity, is  dominated  by  this broad peak at any investigated temperature.

These high-temperature modulations have been named charge density fluctuations (CDF). The name {\itshape fluctuations} is to distinguish CDF from CDW, characterized by a longer correlation length. The Lorentzian peak used to fit the high-temperature signals is indeed very broad. This implies that the correlation length, defined as  $\xi = (\pi \cdot \mathrm{FWHM}_{T_\mathrm{c}})^{-1}$ (here, $\mathrm{FWHM}_{T_\mathrm{c}}$ is the full width at half maximum of the Lorentzian, as determined at the critical temperature $T_\mathrm{c}$) is very short. In particular $2\xi$, representing the diameter of the cluster in which charge order correlation is present, is $\approx 15$ \AA, which is comparable to the modulation period $\lambda = (q_\mathrm{CDW})^{-1} \approx 13$ \AA\, (see Fig. \ref{fig:Fig3}(g)). So, while the CDW correlation diameter is, depending on doping, 3 to 5 times the period of the modulation, the electronic wave of CDF loses its correlation and breaks apart after a single period, which confirms its ultra-short-range nature. 

A second, remarkable, property of these modulations is that they are {\itshape dynamical}, i.e. characterized by a finite energy. To assess this feature, RIXS spectra have been measured using a high energy resolution of 40 meV at  $q_{//} \approx q_\mathrm{CDW}$ and at temperatures above $T_\mathrm{CDW}$, where the CDF contribution to the elastic/quasi elastic region of the spectra is dominant. The pure contribution from the (low-energy) phonons, already minimized by subtracting the analogous spectra measured on the diagonal, should increase upon warming up. On the contrary, the quasi-elastic intensity of the spectra shows a drop when increasing the temperature from 150\,K to 250\,K, which is in agreement with the observed reduction of the CDF intensity at high temperature (see Fig. \ref{fig:Fig3}(h)). The difference between these two spectra, mainly due to the temperature dependence of the CDF, is clearly inelastic (see Fig. \ref{fig:Fig3}(i)). By fitting the difference peak, the energy $\omega_0$ of the charge density fluctuations can be estimated: at the optimally doped regime, it is of the order of 10 meV. In Ref. \citenum{arpaia2019dynamical}, the peak in Figure \ref{fig:Fig3}(i) has been successfully fitted in the framework of the theory of the charge density instability of the high-doping correlated Fermi-liquid, developed for cuprates already in the '90s, and based on the mechanism of frustrated phase separation \cite{castellani1995singular, andergassen2001anomalous, caprara2017dynamical}.

\section{Universality of high-temperature charge modulations in hole-doped cuprates}

The experiment presented in Sec. \ref{sec:sec3} overturns many common beliefs related to charge order: modulations of the electronic density dominate the phase diagram of cuprates, surviving at least up to room temperature, and extending to doping levels beyond those covered by CDW. With this respect, the measurements performed on a slightly overdoped sample ($p \approx 0.19$) confirm the absence of any narrow CDW peak, as already concluded in previous studies \cite{ghiringhelli2012long, blanco2014resonant}, but show at the same time the presence of a broad peak already at low temperatures (see Fig. \ref{fig:Fig3}(c)). CDF are therefore present also in proximity of the putative quantum critical point at $p \approx 0.19$ connected to the pseudogap and in the overdoped regime, disappearing only -- as confirmed by recent STM and X-ray scattering experiments \cite{miao2021charge, li2021evolution} -- in extremely overdoped, non superconducting, samples. Ultimately, the low temperature charge density waves confined to relatively small regions of the cuprate phase diagram appear as atolls in the sea of ultra-short range charge modulations. This finding has  to impact on the description of the physics of HTS, either in the normal or in the superconducting state or in both. 

Many other experiments followed, confirming the presence of high temperature charge modulations in all the hole-doped compounds where CDW were previously observed \cite{yu2020unusual, wang2020doping, lee2021spectroscopic, boschini2021dynamic, miao2017high, miao2019formation, wen2019observation, miao2021charge, lin2020strongly, wang2020high}. In particular, the work by Yu {\itshape et al} \cite{yu2020unusual} found high temperature charge modulations in underdoped HgBa$_2$CuO$_{4+\delta}$ (Hg1201), with  characteristics which are very similar to those presented in Ref. \citenum{arpaia2019dynamical}, although the analysis follows a different strategy. 
\begin{figure}[bpht]
\centering
\includegraphics[width=8cm]{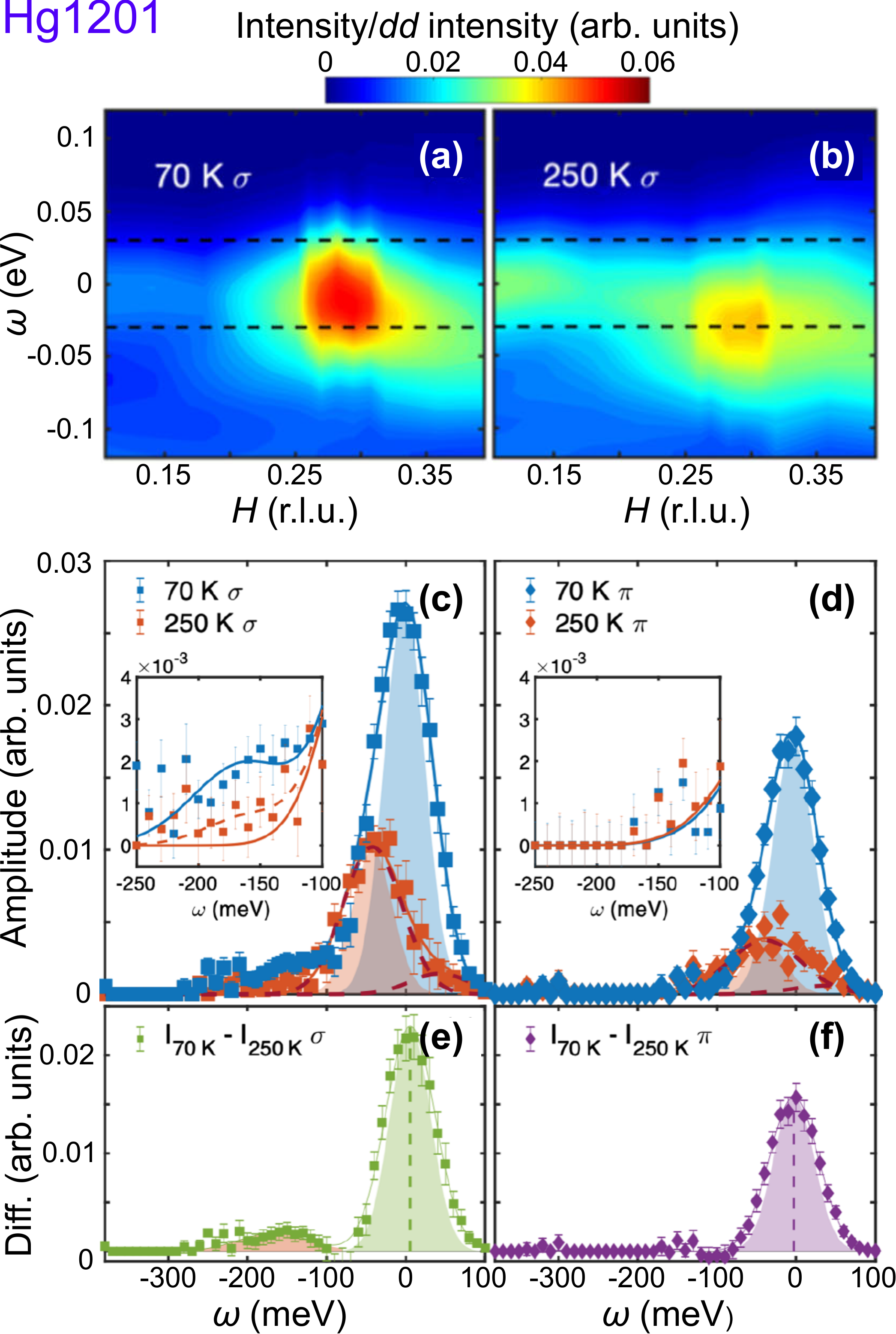}
\caption{{\footnotesize Dynamical Charge Correlations in HgBa$_2$CuO$_{4+\delta}$. (a)-(b) Momentum-energy maps of RIXS spectra, measured with $\sigma$ polarization at 70 and 250\,K on an underdoped ($p$ = 0.086, $T_{\mathrm{c}}$ = 70\,K) sample. The dashed lines show the energy resolution (60 meV) used for the experiment. At high temperature, only a dynamical charge signal, with a characteristic energy of about 40 meV and $q \approx q_{\mathrm{CDW}}$, survives. At both temperatures, the intensity of the RIXS spectra has been integrated at each $q$ in several intervals, whose width equals the energy resolution. The resulting intensity for each interval has been fitted by gaussian peaks, whose amplitude is plotted as a function of energy (c) for $\sigma$ (see squares) and (d) for $\pi$ (see diamonds) polarization. The differences 70\,K - 250\,K of these two data sets are plotted in panels (e) and (f) respectively. { Adapted from Ref. \citenum{yu2020unusual} (\copyright 2020 
The Author(s))}.}} \label{fig:Fig4}
\end{figure}
Here, RIXS spectra are measured  as a function of $q_{//}$ in $\sigma$ polarization at two different temperatures: 70\,K, which is about the critical temperature of the Hg1201 sample, and at 250\,K, which is beyond the onset of quasi-static CDW determined on this compound in previous studies \cite{tabis2014charge}. The momentum-energy maps, focusing on the low-energy region of the RIXS spectra, display at both temperatures a peak centered at $q_{\mathrm{CDW}} \approx 0.28$ r.l.u. (see Figs. \ref{fig:Fig4}(a)-(b)). However, while the peak at low temperature is dominated by the quasi-static component, at high temperature only a dynamical signal survives, at energies $\omega_0 \approx 40$ meV. The electronic nature of the dynamical signal is confirmed by the signal reduction occurring when measuring the spectra in $\pi$ polarization. To isolate the charge order contribution from the background, the intensity of the RIXS spectra has been integrated in several intervals, with width equaling the energy resolution (60 meV). In each interval, the resulting curves are fitted with a Gaussian peak, related to charge order of any kind, superimposed to a polynomial background, mostly due to phonons (at least those varying smoothly with $H$) and scattering from defects and disorder. Finally, the Gaussian intensities are plotted as a function of the energy, so to exclude the background contribution.  This procedure, alternative to the measurement of the ($H$,$H$) scan in Ref. \citenum{arpaia2019dynamical}, and bringing to RIXS spectra dominated by the charge contribution, are reported in Figures \ref{fig:Fig4}(c)-(d) (see squares and diamonds) for both investigated temperatures and polarizations. The spectra are characterized by three distinct charge features, whose intensity significantly drops in $\pi$ polarization: the first one, quasi-static, is related to CDW; the second one, dynamical at $\omega_0 \approx 40$ meV, is due to CDF; the third one, still dynamical, is peaked at $\sim$ 165 meV, in the energy scale of magnetic excitations. From the difference 70\,K - 250\,K (see Figs. \ref{fig:Fig4}(e)-(f)), the temperature dependence of these features can be estimated: the CDW decrease when increasing temperature, disappearing at 250\,K. The dynamical high-energy charge correlations show the same behavior, but they persist, although with a smaller intensity, also at 250\,K; finally, CDF are nearly temperature independent, as already observed in Ref. \citenum{arpaia2019dynamical}. In addition to that the analysis of Figure \ref{fig:Fig4}(c) shows that the CDF peak is not only broad in $q$, but also broad in energy, since the FWHM of the peak characterizing them is larger than the instrumental energy resolution. 

Once the presence and the main characteristics of these high temperature charge modulations have been properly assessed, at least four main questions arise. They are currently motivating new investigations, but some preliminary answers, both theoretical and experimental, are presented in this and the two following sections. 
\begin{enumerate}
  \item What is the link between quasi-static CDW and dynamical CDF?
  \item What are the possible implications of CDF for the physics of cuprates?
  \item  What is the coupling -- if  any -- between dynamical CDF and other low-energy (i.e. lattice and spin) excitations?
 \item  Are the characteristics of high temperature charge modulations universal among the different cuprate families?
  \end{enumerate}

Regarding the first question, it might be tempting to consider CDF as a mere continuation of CDW at high temperatures. In this view, $T_{\mathrm{CDW}}$ would have been strongly underestimated in previous RXS studies, and high-$T$ charge modulations would have appeared now just because of a better sensitivity of the RIXS facilities. However, many experimental evidences show this is not the case, even though $T_{\mathrm{CDW}}$, as measured in recent RIXS studies \cite{arpaia2019dynamical, wahlberg2020restoring} and by using other techniques \cite{bakr2013lattice}, looks higher than in the pioneering experiments. First of all, the $q$ vectors of CDW and CDF are very similar, which is an indication of their common nature, but not identical. Additionally, CDF have a fluctuating nature, whereas CDW emerge at lower temperature as a nearly static, quasi-critical modulations, associated to the divergence of a correlation length at a quantum critical point. In particular, the nearly temperature independence of the CDF intensity and correlation length has several consequences: on one side, it shows that CDF are not competing with the superconducting order, differently than CDW, whose intensity and correlation length significantly drop when entering in the superconducting state; on the other side, it implies that CDF are still present at low temperatures, coexisting with CDW. A possible picture compatible with the latter occurrence is that only a small fraction of CDF evolves into CDW when decreasing the temperature. A crucial role would be played in that case by disorder as theoretically predicted \cite{nie2014quenched, capati2015electronic}, and experimentally investigated by scanning tunneling microscopy \cite{pan2001microscopic, kohsaka2004imaging} or X-ray scattering \cite{achkar2014impact, campi2015inhomogeneity} techniques, which prevents the development of charge modulations with longer correlation length when the sample is cooled down. Finally, the response to strain is very different for CDW and CDF. The in-plane uniaxial compression (expansion) along one crystallographic axis strongly modifies charge density waves, giving rise to their enhancement (suppression) in the orthogonal direction; on the contrary, no appreciable effects of strain have been detected for charge density fluctuations  \cite{wahlberg2020restoring, boyle2020large}. Thus, even though low temperature CDW and high temperature CDF seem to have a common nature, and CDW are possibly deriving from CDF, they are two distinct features, presenting peculiar characteristics, and can therefore affect differently the physics of cuprate HTS. 

In this respect, coming to the second open question, it is appealing to verify to what extent the newly-discovered, very short-range, charge modulations determine the phenomenology of HTS, either in the normal or in the superconducting state, since they are present in such broad ranges of temperature and doping and are characterized by finite energies in the order of meV. This energy scale is comparable to both the pseudogap and the superconducting gap, which might suggest an intertwining between these two phenomena and charge order, in agreement with recent Raman experiments on HgBa$_2$Ca$_2$Cu$_3$O$_{8+\delta}$ and YBa$_2$Cu$_3$O$_{7-\delta}$ \cite{loret2019intimate, loret2020universal}.
In Ref. \citenum{yu2020unusual},  the possible contribution of the charge modes to the superconducting state is investigated. The two dynamical charge contributions and the paramagnons determined by RIXS on Hg1201 are indeed characterized by energies comparable to those of two main peaks of the pairing glue function, as determined on the same compound by optical spectroscopy measurements \cite{van2009optical}. This would be consistent with previous theoretical proposals, explaining the $d$-wave pairing in HTS in terms of incommensurate charge order, in combination with magnetic excitations \cite{castellani1996non, wang2015enhancement}.
Moving instead to the normal state, the high-energy charge feature singled out in Ref. \citenum{yu2020unusual} at $\sim$165 meV has been thereby discussed as an imprint of the pseudogap, characterized by a similar energy value in several optimally doped cuprates, as measured by ARPES, optical conductivity and tunneling experiments \cite{honma2008unified}. However, dynamical charge density fluctuations are among the few excitations surviving up to temperatures well beyond the pseudogap, where the strange metal region dominates the cuprate phase diagram: different theoretical proposals have therefore associated CDF to the phenomenology occurring in that region, whose main benchmark -- as observed in the transport measurements -- is represented by the linear behaviour of the electrical resistivity $\rho$ as a function of the temperature $T$, from the pseudogap temperature up to the highest investigated temperatures \cite{ando2004electronic, barivsic2013universal, arpaia2018probing}. Such property is based on an isotropic scattering rate, affecting equally all the states on the Fermi surface. Charge density fluctuations are an appealing candidate to mediate such isotropic scattering, since they produce a broad peak in the reciprocal space.  Moving from this point, Seibold {\itshape et al} \cite{seibold2021strange} calculate the resistivity $\rho$ within a classic Boltzmann approach, using the quasi-particle scattering rate along the Fermi surface due to the presence of charge density fluctuations: the linear-in-$T$ behavior of $\rho$ is found, and the $\rho (T)$ of optimally doped and slightly overdoped (Y,Nd)Ba$_2$Cu$_3$O$_{7-\delta}$ is quantitatively fitted from room temperature respectively down to $T^*$ and down to $T_{c}$. Interestingly, the presence of high temperature charge fluctuations is instrumental for building up the strange metal state even for holographic quantum matter theories. Here, the linear-in-$T$ behavior of the resistivity is explained in terms of a new fundamental time scale, defined by the Planckian time $\tau = \hbar/k_BT$ (where $\hbar$ and $k_B$ are the reduced Planck and the Boltzmann constants), representing a lower bound on the  time  for  a  system  to  reach  local  equilibrium  as $T \xrightarrow{} 0$ \cite{delacretaz2017bad, amoretti2019universal, amoretti2020hydrodynamical}. Within this framework, fluctuating density waves, breaking the translational symmetry, are needed to explain the temperature dependence of the optical conductivity of bad metals, with the Drude peak which broadens and moves to non-zero frequencies $\omega \propto 1/\tau$ as the temperature increases \cite{tsvetkov1997plane, osafune1999pseudogap, takenaka2003incoherent}. It is therefore remarkable that theories so different among each other bring to the same conclusion: charge density fluctuations are likely the long-sought microscopic mechanism underlying the peculiarities of the metallic state of cuprates.

The third question is related to the possible coupling between dynamical charge fluctuations and other, low-energy excitations, as phonons and magnons. In particular, the nearly $T$-independent signal observed in Refs. \citenum{arpaia2019dynamical, yu2020unusual} is characterized by energies of tens of meV, which are in the same order as several optical phonon branches. This could possibly enhance the intertwining of charge and lattice modes, which might have a decisive  role for the HTS phenomenology. Indeed, optical spectroscopy measurements with simultaneous time and frequency resolution on optimally doped Bi2212 crystals have shown that electronic and lattice processes coexist in the bosonic excitations which are supposed to mediate the formation of Cooper pairs \cite{dal2012disentangling}. An influence on the crystal lattice, given by the spatial arrangement of the electronic cloud, was easy to expect, and it was indeed confirmed by several, rather old, inelastic X-ray and neutron scattering (respectively, IXS and INS) measurements. The phonon dispersion was exhibiting a pronounced softening, mainly involving the bond stretching branches, in several superconducting hole-doped and electron-doped cuprates, as YBCO \cite{chung2003plane, pintschovius2004oxygen}, Bi2201 \cite{graf2008bond}, Hg1201 \cite{uchiyama2004softening}, LSCO \cite{mcqueeney1999anomalous, pintschovius1999anomalous} LBCO \cite{reznik2006electron} and NCCO \cite{d2002anomalous, braden2005dispersion}. This softening could not be explained by simple shell models, and required to postulate the presence of spatial charge fluctuations. Later on, the wave vector $q$, at which the softening observed with IXS and INS was maximum, was discovered to be very close to the $q_{\mathrm{CDW}}$ determined by RXS \cite{ghiringhelli2012long, comin2014charge, da2014ubiquitous}. Finally, the recent advances in the RIXS technique, with its improved energy resolution, has given the opportunity to study phonons, isolating their contribution from the pure elastic/quasi elastic intensity and even distinguishing between different phonon branches \cite{rossi2019experimental, braicovich2020determining, peng2020enhanced}. As a consequence, it is nowadays possible to study at the same time and possibly separate charge excitations and phonon softening. The first study of such kind is the report by Chaix {\itshape et al} \cite{chaix2017dispersive}. 
\begin{figure}[b]
\centering
\includegraphics[width=8cm]{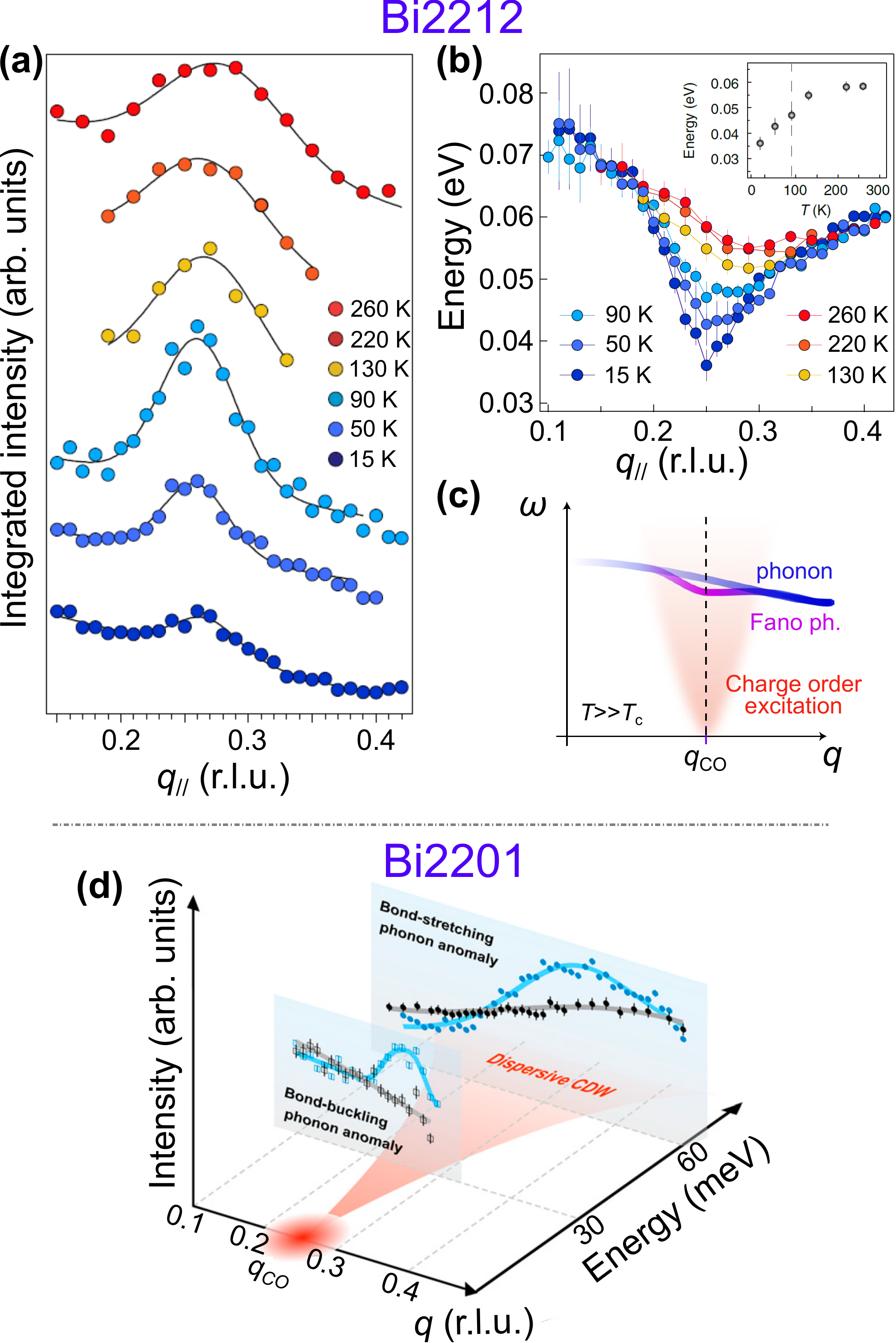}
\caption{{\footnotesize Intertwining of charge excitations and phonons in Bi-compounds. (a) A charge order peak is obtained at every temperature in nearly optimally doped  Bi$_2$Sr$_2$CaCu$_2$O$_{8+\delta}$, by integrating the quasi-elastic region of the RIXS spectra. At low energies, the spectra can be fitted by considering a quasi-elastic peak and a higher energy peak due to bond-stretching phonons. The fitted phonon dispersion is presented in panel (b) as a function of the temperature. At high temperatures, the phonon softening still occurs, although smaller. The sketch of the model describing the physics into play is shown in panel (c): with respect to low temperatures, the charge order excitation continuum is still present, but it is more damped in energy and broadened in momentum. As a consequence,  the Fano effect is weaker. { Panels (a)-(b)-(c) are adapted from Ref. \citenum{lee2020spectroscopic}. (\copyright 2020 
Wei-Sheng Lee)}. (d) A similar study is done on underdoped Bi$_2$Sr$_{1.4}$La$_{0.6}$CuO$_{6+\delta}$. Here, the low-energy region of the spectra show, in conjunction with CDW, an anomaly of the bond-buckling phonons in addition to that of the bond-stretching phonons (see blue circles and squares). Again, the effect - sketched in the panel - is explained in terms of phonon interplay with dynamical charge order excitations. {Adapted from Ref. \citenum{li2020multiorbital} (\copyright 2020 
National Academy of Sciences)}.}} \label{fig:Fig5}
\end{figure}
Here, a gradual softening of the bond-stretching phonons is observed at 20\,K in underdoped ($p \approx 0.08-0.09$) Bi2212, with a minimum energy of 40–45\,meV near $q_{\mathrm{CDW}}$. The softening is accompanied to an enhancement of the phonon intensity, peaked at $q_{\mathrm{a}} = q_{\mathrm{CDW}} + 0.07$ r.l.u.. Since the RIXS phonon cross-section is directly connected to the strength of the electron–phonon coupling at a given momentum \cite{ament2011determining, devereaux2016directly, braicovich2020determining, geondzhian2020generalization}, these anomalies can be explained by considering a strong interference (i.e. the Fano effect) between the phonon branch and the underlying, dispersive charge density excitations. These excitations present a funnel-like spectral weight, emanating from $q_{\mathrm{CDW}}$ and extending up to $\sim$ 60\,meV. More recently, a careful temperature dependence of this interference effect has been performed, in nearly optimally doped Bi2212, in the work by Lee {\itshape et al} \cite{lee2021spectroscopic}.
The quasi-elastic region of the RIXS spectra is fitted by considering a lower energy peak due to charge order and a higher energy peak due to bond stretching  phonons. The charge order peak (see Fig. \ref{fig:Fig5}(a)), as well as the phonon softening (see Fig. \ref{fig:Fig5}(b)), survive up to the highest investigated temperature (260\,K), well beyond the pseudogap temperature.
However, while the charge order peak is strongly suppressed below $T_{\mathrm{c}}$, due to the competition between CDW and superconductivity, the phonon softening -- occurring already at 260\,K -- increases as the temperature is lowered, reaching its maximum at 15\,K, where the phonon energy drops down to $\sim$ 35\,meV near the charge order wavevector $q_{\mathrm{CO}}$. The different temperature dependence of CDW and phonon softening shows that the Fano effect cannot be directly due to the presence of charge density waves.  Viceversa, a continuum of charge order-driven, inelastic quantum fluctuations is postulated, whose intensity is maximum in correspondence of the putative quantum critical point (QCP) at $p \approx 0.19$, while it decreases when increasing the dissipation (i.e. the temperature) and moving away from the QCP doping level. So, within this framework, the high-temperature measurements ($T{\gg}T_{\mathrm{c}}$) can be described considering the interference between the funnel-like continuum of dynamical charge order fluctuations originating at $q_{\mathrm{CO}}$, still present although less intense than at lower temperatures, and the bond stretching phonons, giving rise to a weak phonon softening  (see Fig. \ref{fig:Fig5}(c)). The effect of dynamical charge fluctuations on phonons is further investigated in the work by Li {\itshape et al} \cite{li2020multiorbital}. Here, RIXS spectra are measured both at the Cu $L_3$ and at the O $K$ edges on the single-layer, underdoped, Bi2201. At the Cu $L_3$ edge, with an energy resolution of 40 meV, the same analysis of Ref. \citenum{lee2021spectroscopic} is performed, showing the presence of quasi-static CDW at $q_{\mathrm{CO}}$ and the softening of the bond stretching phonons, whose intensity is enhanced at $q_1>q_{\mathrm{CO}}$ (see blue circles in Fig. \ref{fig:Fig5}(d)). At the O $K$ edge the energy resolution of 26 meV allows to fit the low-energy region of the spectra  with three peaks, isolating the bond buckling branches at energies $\sim$30\,meV from the bond stretching branches at energies $\sim$60\,meV. The result is that the funnel-like continuum of dynamical charge fluctuations originating at $q_{\mathrm{CO}}$ interferes also with the bond-buckling modes, giving rise to the Fano effect and to a peak of the phonon intensity centered at $q_2>q_{\mathrm{CO}}$ (see blue squares in Fig. \ref{fig:Fig5}(d)). The same analysis, performed on an overdoped Bi2201 presenting no signature of charge order, shows the disappearance of any phonon anomaly (see black symbols in Fig. \ref{fig:Fig5}(d)). These results confirm therefore the strong intertwining occurring between charge and lattice excitations, resulting in interference effects covering a wide range of energies and phonon branches.

Finally, the detection of additional charge fluctuations even at energies of hundreds of meV leads to the issue of a possible coupling between charge and magnetic excitations. In the previously discussed Ref. \citenum{yu2020unusual}, the energy and damping of the paramagnons are unaffected by dynamical charge fluctuations, which would bring to exclude any sort of coupling. In another recent work, Boschini  {\itshape et al} \cite{boschini2021dynamic} investigate the charge order phenomenon in Bi2212 at different oxygen doping, combining EI-RXS and low energy resolution (0.8 eV) RIXS. In addition to the quasi-static/low energy charge order signal, centered at $q_{\mathrm{CDW}}$ along the two crystallographic axis and surviving up to the highest investigated temperature ($T = 360$\,K), they see a dynamical signal originating in the energy loss range between 500 and 900 meV. The main characteristic of this signal is that, although centered at $q_{\mathrm{CDW}}$, it shows a ring-like structure in the $q_{\mathrm{x}}/q_{\mathrm{y}}$ plane. Such geometry would point toward a strong connection between charge order and the nematic, rotational  symmetry breaking order, accounting for the very different nematic states discovered in cuprates so far \cite{wu2017spontaneous, sato2017thermodynamic, murayama2019diagonal}.  Higher energy resolution measurements, with minimized cross-talk effects between different region of the spectra and well defined integration energy ranges, will be needed in order to get a better understanding about the nature of this ring-like, dynamical, charge order and about its possible connection with magnetic excitations.

More in general, evidences of entwining between charge and magnetic correlations have not been found in cuprates so far, with the only important exception of the 214 La-based family, where charge and spin are locked, giving rise to the ``stripe'' order. One can expect that the peculiarities of these compounds get reflected also on the properties of the  high temperature charge modulations. Our fourth open question is indeed about the universality of these newly-discovered signals, and it will be the matter of the next two sections.

\section{High-temperature precursors of stripes in the 214 family}

The low-temperature charge order phenomenon in the 214 family of cuprates is clearly distinct from that in the other families of hole-doped cuprates \cite{tranquada1995evidence, fujita2004stripe, croft2014charge}. Here, the dominant spin degree of freedom makes CDW strongly intertwined with the spin density wave order. In particular, at the $p = 1/8$ doping, where this ``stripe'' order is strongest, the CDW wave vector is almost commensurate to a period of 4 unit cells, and it is related to the SDW wave vector $\delta_{\mathrm{SDW}}$ by the relation $q_{\mathrm{CDW}} = 2\delta_{\mathrm{SDW}}$. Finally, the  CDW  wave vector increases (weakly)  with  doping, rather than decreasing. So, while the tendency to charge ordering seems universal among cuprates, the peculiarities of these charge modulations are not.

\begin{figure*}[ht]
\centering
\includegraphics[width=17cm]{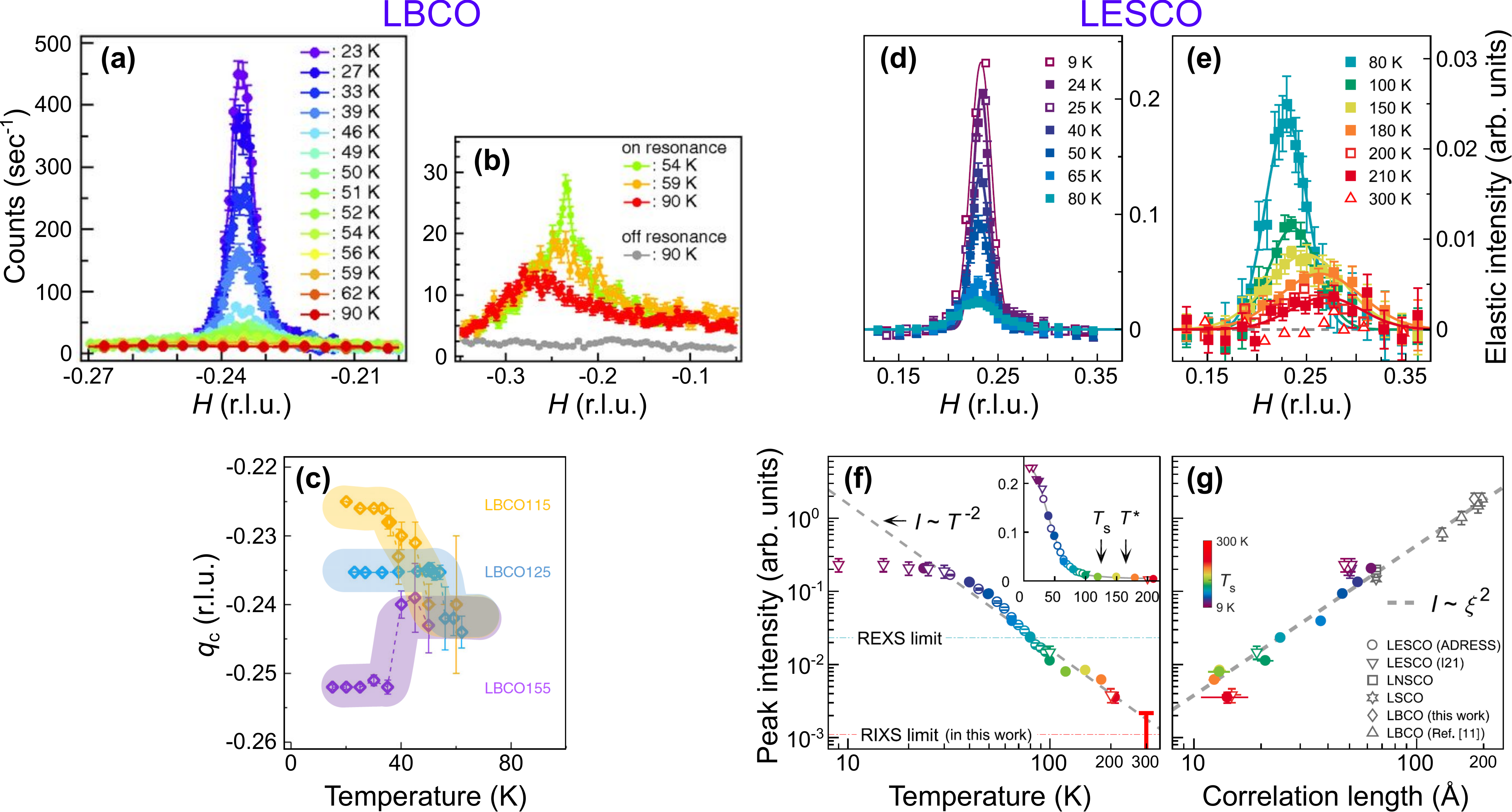}
\caption{{\footnotesize Stripes and stripe precursors measured with RIXS in La-based cuprates. (a)-(b) Temperature dependence of charge order in the stripe-ordered cuprate La$_{1.875}$Ba$_{0.125}$CuO$_4$. At temperatures above the onset of the spin-charge locking, connected to the stripe order, a broad-in-$q$ peak persists, although with a different wavevector $q_{\mathrm{c}}$. {Adapted from Ref. \citenum{miao2017high} (\copyright 2017 
National Academy of Sciences)}. (c) Performing the same temperature investigation on several La$_{2-x}$Ba$_{x}$CuO$_4$ samples with different doping level, it has been discovered that the wavevector of the short-ranged precursors of the stripes at high temperatures is doping-independent and quasi-commensurate. { Adapted from Ref. \citenum{miao2019formation} (\copyright 2019 
The Author(s))}. (d)-(e) Temperature dependence of charge-stripe order in La$_{1.675}$Eu$_{0.2}$Sr$_{0.125}$CuO$_4$: short range correlations survive at temperature beyond the pseudogap. {Adapted from Ref. \citenum{wang2020high} (\copyright 2020 
American Physical Society)}. (f)-(g) The intensity $I$ of the charge-stripe peak decays vs temperature as $T^{-2}$, while the correlation length $\xi$ as $T^{-1}$. As a consequence the volume of the peak, proportional to $I\cdot\xi^{-2}$, remains roughly temperature independent.  {Adapted from Ref. \citenum{wang2020high} (\copyright 2020 
American Physical Society)}.}} \label{fig:Fig6}
\end{figure*}

Spin fluctuations were known to survive above the onset temperature of stripe order. However, charge fluctuations remained undetected until the improved sensitivity of RIXS instrumentation started to be exploited. The discovery of CDW correlations, existing at temperatures  above  the  onset $T_{\mathrm{CDW}}$, was done in Ref. \citenum{miao2017high},  where the  canonical stripe-ordered cuprate La$_{1.875}$Ba$_{0.125}$CuO$_4$ ($p = 1/8$) was explored up to $T = 90$\,K. Integrating the quasi-elastic intensity of the RIXS spectra, a narrow peak can be observed at the lowest temperatures, corresponding to the well-known CDW with a wavevector $q_{\mathrm{CDW}} = 0.235$ r.l.u. and a rather long correlation length. When increasing the temperature, the intensity of the CDW peak drops and seems to disappear around 55\,K, onset of the charge stripe order as determined by previous x-ray and neutron diffraction experiments  \cite{hucker2011stripe} (see Fig. \ref{fig:Fig6}(a)). However, above this temperature a weak, broad-in-$q$ peak persists, which is the signature of short-ranged charge fluctuations (see Fig. \ref{fig:Fig6}(b)). These modulations, whose hints are present already at base temperature, survive up to the highest investigated temperature and are characterized by a wave vector shifting to values higher than $q_{\mathrm{CDW}}$. This last occurrence is crucial, since it highlights that the relation $q_{\mathrm{CDW}} = 2\delta_{\mathrm{SDW}}$ is violated: high-temperature charge fluctuations decouple from spin fluctuations, whose wave vector presents a different temperature dependence above the CDW onset temperature \cite{fujita2004stripe}. In a more recent report, Miao {\itshape et al} \cite{miao2019formation} studied with RIXS both the temperature and the doping dependence of the charge order phenomenon in LBCO. They discovered that, while CDW occur at different wave vectors depending on the doping, the high temperature charge fluctuations - precursors of the stripes - are doping independent and characterized by a quasi-commensurate wave vector which emerges at $T>T_{\mathrm{CDW}}$ (see Fig. \ref{fig:Fig6}(c)). The possible dynamical nature of these high-temperature modulations is discussed in the article, as well as the coupling to the phonon branches. 

High temperature charge modulations have been also measured in La$_{2-x}$Sr$_{x}$CuO$_{4}$ over an extensive doping range, including the slightly overdoped level ($p=0.21$), where they precede striped CDW, coexisting with them below $T_{\mathrm{CDW}}$ \cite{lin2020strongly, miao2021charge}. Both CDW and their precursors disappear instead when the doping is increased even further ($p=0.25$) and the samples, weakly superconducting ($T_{\mathrm{c}} = 10$\,K), are characterized by a Fermi-liquid behavior in the whole normal state. 

Wang {\itshape et al} have recently investigated the charge-stripe order in underdoped ($p = 1/8$) La$_{1.675}$Eu$_{0.2}$Sr$_{0.125}$CuO$_4$ in a very broad temperature range \cite{wang2020high}. The charge order peak evolves continuously from base temperature, where it is intense and narrow, up to room temperature, where it is weak and broad (see Figs. \ref{fig:Fig6}(d)-(e)). The charge-stripe peak extends therefore beyond the structural low-temperature tetragonal phase, whose onset is at $\approx 125$\,K \cite{achkar2016nematicity}, and beyond the pseudogap temperature $T^* \approx 160$\,K, determined by transport measurements \cite{cyr2018pseudogap}. Even though the wave vector of the modulation is locked at low temperatures and shifts upon warming up the sample, the peak is here analyzed following a different approach with respect to the previous works on La-based cuprates. Indeed, the peak is interpreted in terms of a single component, whose onset $T_{\mathrm{CDW}}$ is above the highest investigated temperature. This is because the intensity of the peak decays, with the exception of the lowest temperatures, as $T^{-2}$, while its correlation length as $T^{-1}$ (see Figs. \ref{fig:Fig6}(f)-(g)): the volume of the peak, considered in its whole, and given by $I\cdot\xi^{-2}$, remains therefore roughly temperature independent. The much lower $T_{\mathrm{CDW}} \approx 80$\,K measured in previous studies \cite{fink2009charge, achkar2016nematicity} is attributed to the lower sensitivity of the EI-RXS instruments with respect to RIXS. Finally, the dependence of the charge scattering intensity on the correlation length (see Fig. \ref{fig:Fig6}(g)) looks as a general property of the different La-based compounds, reminiscent of the dynamic magnetic critical scattering in La$_2$CuO$_4$ \cite{birgeneau1999instantaneous}, and associated with two-dimensionality and local spin nature.

Because of the instrumental energy resolution, the dynamical nature of the charge-striped peak at high temperature has not been discussed in the aforementioned work. However, it has been predicted that high temperature charge correlations in the 214 family are dynamical \cite{kivelson2003detect, reznik2006electron}, and hints of such nature have been obtained by transport \cite{li2007two} and pump-probe \cite{torchinsky2013fluctuating} measurements. In this respect, a recent report by Mitrano {\itshape et al} reveals dynamical charge order in LBCO by time-resolved resonant soft X-ray scattering at a free-electron laser \cite{mitrano2019ultrafast}. Fluctuations at picosecond time scales, propagating by Brownian-like diffusion, are characterized by energies $\omega_0$ close to the superconducting energy $k_{\mathrm{B}}T_{\mathrm{c}}$.

In conclusion, the peculiarities of CDW in La-based compounds, listed in the beginning of this section, are mitigated at high temperatures: there, charge modulations, precursors of the low temperature stripes, still persist, but they do not present any locking to spin excitations and have a possible dynamical character. These characteristics appear therefore universal among all the hole doped families.  

\section{Intertwining of charge order and magnetic fluctuations: the case of electron-doped cuprates}

Finally, we move the focus to electron doped cuprates, to find out whether high temperature charge modulations have been detected in these compounds  and, in the affirmative case, which are their characteristics.

\begin{figure}[htbp]
\centering
\includegraphics[width=8cm]{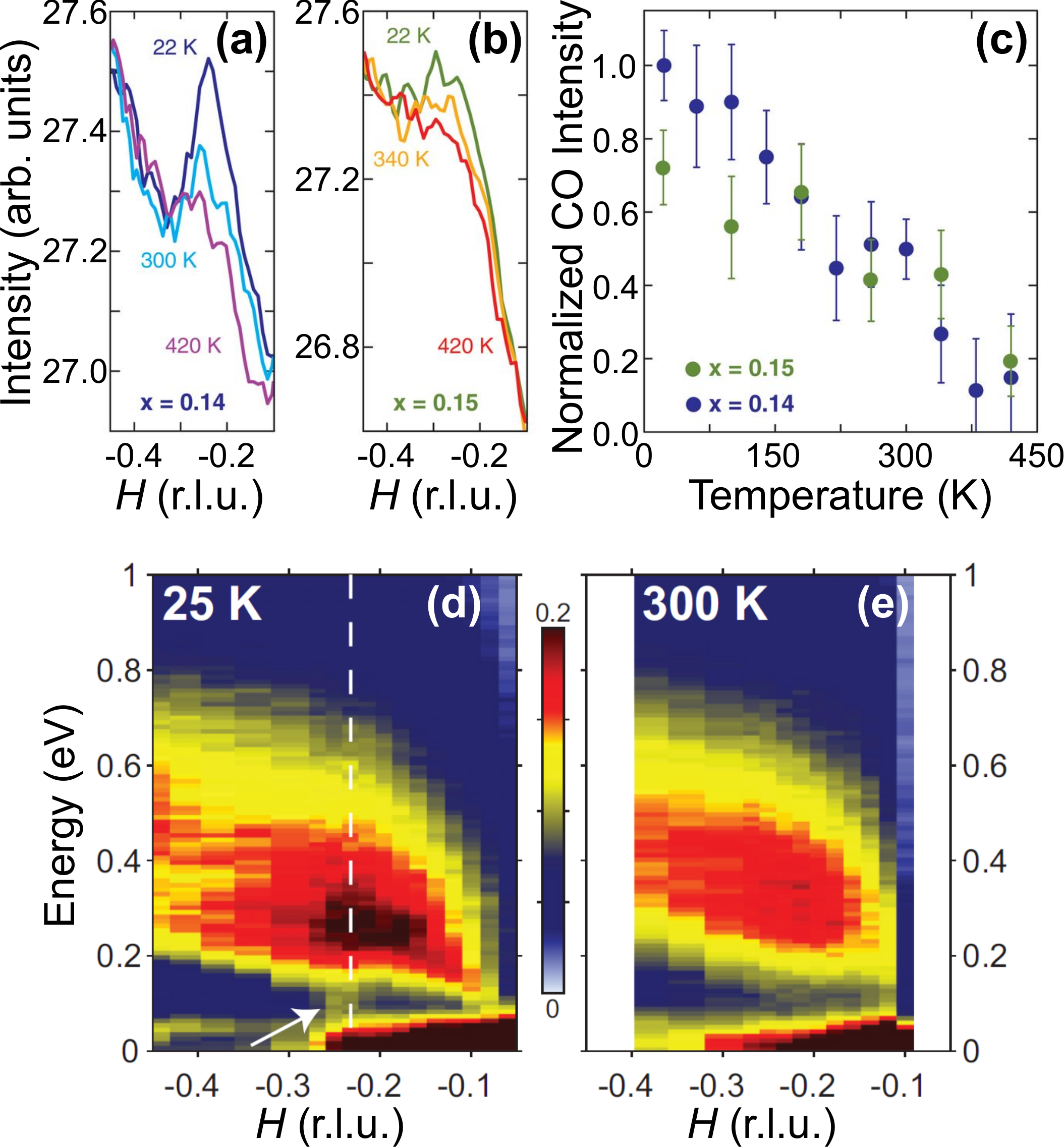}
\caption{{\footnotesize  Dynamical charge correlations in Nd$_{2-x}$Ce$_x$CuO$_4$. (a)-(b) Energy-integrated RXS measurements on two samples with different doping level show a temperature dependent charge peak centered at $q_{\mathrm{c}}$, whose onset is above room temperature. {Adapted from Ref. \citenum{da2015charge} (\copyright 2015 
American Association for the Advancement of Science)}. (c) Temperature dependence of the RXS intensity for the two samples in panels (a) and (b), obtained from the maxima of the background-subtracted peaks. { Adapted from Ref. \citenum{da2015charge} (\copyright 2015 
American Association for the Advancement of Science)}. (d)-(e) Momentum-energy maps of RIXS spectra, measured with $\sigma$ polarization at 25 and 300\,K on a non superconducting sample with $x = 0.106$. The temperature dependent signal, as in panels (a)-(b), mostly originates from dynamical charge correlations at $q_{\mathrm{c}}$, occurring in the same energy range spanned by the magnetic excitations. { Adapted from Ref. \citenum{da2018coupling} (\copyright 2018 
American Physical Society)}.}} \label{fig:Fig7}
\end{figure}

In one of the pioneering works, da Silva Neto {\itshape et al} investigate two NCCO samples with electron doping $n = 0.14$ and $0.15$ using EI-RXS \cite{da2015charge}. In both samples, they reveal a charge order signal, which resonates at the Cu $L_3$ absorption edge. The peak, centered at a wavevector $q_{\mathrm{CO}} \approx 0.24$ r.l.u., continuously drops its intensity when increasing the temperature, indicating that the onset temperature is above the upper investigated temperature of 420\,K (see Figs. \ref{fig:Fig7}(a)-(b)-(c)). 
{Thereafter, a peak with a similar temperature dependence has been measured by EI-RXS also in non superconducting NCCO samples \cite{da2016doping, jang2017superconductivity}.} To get more insights into this peak and determine whether its nature is elastic or inelastic, a RIXS experiment was later performed on NCCO \cite{da2018coupling}. While decomposing the RIXS spectra in several energy windows, the authors conclude that at low temperature only half of the intensity measured at $q_{\mathrm{CO}}$ comes from the quasi-elastic region (defined for energies lower than 60 meV). The remaining intensity derives, as evident in the energy momentum maps plotted in Figure \ref{fig:Fig7}(d), from a strongly dynamical peak, centered at $q_{\mathrm{CO}}$ but having energies comparable to those spanned by paramagnons. Upon warming the sample to room temperature, the quasi elastic peak disappears, while the dynamical peak persists although with decreasing intensity  (see Fig. \ref{fig:Fig7}(e)). The phenomenology of this high energy peak, related to charge correlations, resembles that of the analogous peaks recently measured in Hg1201 \cite{yu2020unusual} and Bi2212 \cite{boschini2021dynamic}. However, for the hole doped cuprates a possible intertwining with magnetic excitations was excluded, or at least not discussed. Here, instead, the picture looks clearer. By measuring spectra with a newly developed polarimeter \cite{braicovich2014simultaneous} and therefore resolving the polarization of the scattered photons, the authors find that the dynamic peak centered at $q_{\mathrm{CO}}$ is mostly due to spin-flip processes: this occurrence highlights  a direct coupling between dynamic charge order correlations and magnetic excitations \cite{da2018coupling}.

A similar analysis, with control of the polarization of the scattered photons, still has to be performed on hole-doped cuprates to determine whether this coupling is a peculiarity of electron-doped cuprates or, vice versa, is a general property of charge correlations extending up to high temperatures.

\section{Conclusions and final remarks}
\begin{figure}[b]
\centering
\includegraphics[width=7cm]{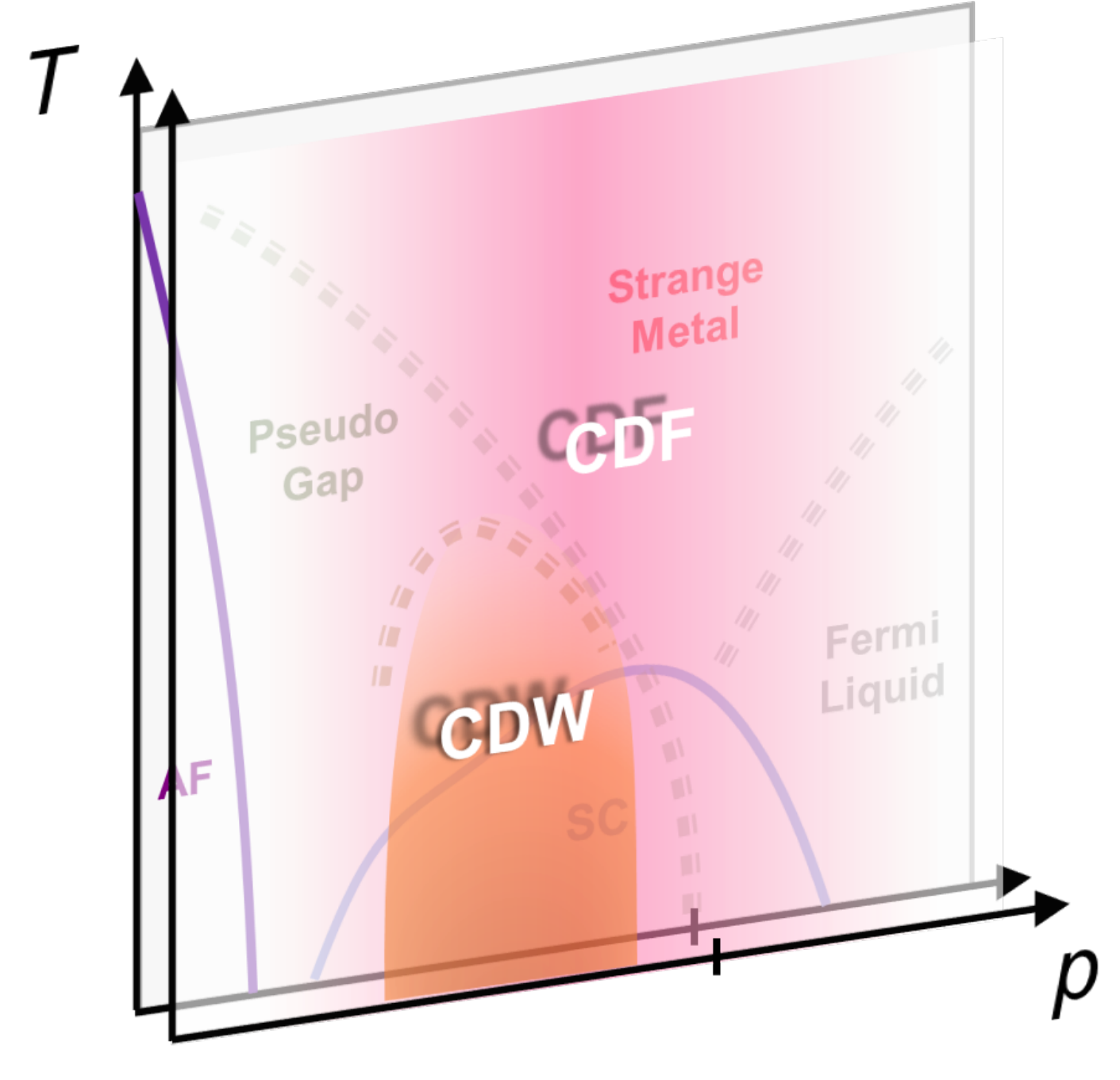}
\caption{{\footnotesize The revised temperature vs doping phase diagram of cuprates. After the discovery of high temperature charge density fluctuations, in addition to the already well-known charge density
waves, charge order dominates the phase diagram both in temperature and doping.}} \label{fig:Fig8}
\end{figure}

The experiments described in this review, most of which performed in the last three years, have shown the presence of charge density fluctuations in broad ranges of doping and temperature (see { Table \ref{tab1}} and Fig. \ref{fig:Fig8}). 
\begin{table*}[thb]
\caption{The most relevant parameters of the high-temperature charge density modulations, recently revealed by RIXS, are here listed for several hole doped and electron doped HTS cuprate compounds. The first two columns indicate the region, in the temperature-doping phase diagram, where the modulations have been detected so far. Third and fourth columns show instead the wavevector and the FWHM of the peak in momentum space, signature of the charge modulations. For each compound, both these quantities are listed as measured at the highest investigated temperature, in order to minimize the effect of the low temperature CDW. Finally, the energies of the charge modulations are listed, as determined by the peak(s) in the energy space, measured at the charge order wavevector: fifth and sixth columns show the energy in the range respectively of lattice and magnetic excitations. The energy is defined ``quasi-elastic'' when the static/dynamical character of the high temperature charge modulations still has to be completely assessed.}
\label{tab1}
\begin{center}
\begin{tabular}{ccccccc}
\hline
\multicolumn{1}{p{1.6cm}}{\centering Compound} & 
\multicolumn{1}{p{1.7cm}}{\centering Onset \\ Temperature \\ (K)} &
\multicolumn{1}{p{3cm}}{\centering Doping \\ Level} &
\multicolumn{1}{p{2.7cm}}{\centering High-$T$ \\ Wavevector \\ (r.l.u.)} &
\multicolumn{1}{p{1.9cm}}{\centering High-$T$ \\ FWHM \\ (r.l.u.)} &
\multicolumn{1}{p{3.4cm}}{\centering Energy \\ (meV)} &
\multicolumn{1}{p{2cm}}{\centering High Energy \\ feature \\ (meV)} \\
\hline
YBCO & $>280$ \cite{arpaia2019dynamical} & 0.11--0.18 \cite{arpaia2019dynamical} & 0.29--0.32 \cite{arpaia2019dynamical} & 0.15--0.18 \cite{arpaia2019dynamical} & $\sim 10$ \cite{arpaia2019dynamical}  & --- \\
Hg1201 & $>250$ \cite{yu2020unusual} & 0.09 \cite{yu2020unusual} & $\sim 0.28$ \cite{yu2020unusual} & --- & $\sim 40$ \cite{yu2020unusual}  & $\sim 165$ \cite{yu2020unusual} \\
Bi2212 & $>300$ \cite{boschini2021dynamic} & 0.09--0.17 \cite{chaix2017dispersive,lee2021spectroscopic,boschini2021dynamic} & 0.25--0.30 \cite{chaix2017dispersive, lee2021spectroscopic, boschini2021dynamic} & $\sim 0.09$ \cite{lee2021spectroscopic} & quasi-elastic \cite{chaix2017dispersive, lee2021spectroscopic, boschini2021dynamic}  & 100--1000 \cite{boschini2021dynamic} \\
Bi2201 & $>300$ \cite{boschini2021dynamic} & 0.10 \cite{boschini2021dynamic} & $\sim 0.29$ \cite{boschini2021dynamic} & --- & quasi-elastic \cite{boschini2021dynamic}  & --- \\
LSCO & $>100$ \cite{miao2021charge} & 0.12--0.21 \cite{miao2021charge} & 0.23--0.24 \cite{miao2021charge} & 0.06--0.18 \cite{miao2021charge} & quasi-elastic \cite{miao2021charge}  & --- \\
LBCO & $>90$ \cite{miao2017high, miao2019formation} & 0.12--0.16 \cite{miao2017high, miao2019formation} & $\sim 0.24$ \cite{miao2019formation} & 0.06--0.10 \cite{miao2019formation} & quasi-elastic \cite{miao2017high, miao2019formation}  & --- \\
LESCO & $>210$ \cite{wang2020high} & 0.125 \cite{wang2020high} & $\sim 0.27$ \cite{wang2020high} & $\sim 0.10$ \cite{wang2020high} & quasi-elastic \cite{wang2020high}  & --- \\
NCCO & $>420$ \cite{da2015charge} & 0.11--0.17 \cite{da2015charge, da2016doping, jang2017superconductivity, da2018coupling} & 0.16--0.28 \cite{da2016doping} & $\sim 0.13$ \cite{jang2017superconductivity} & quasi-elastic \cite{da2018coupling}  & $\sim 250$ \cite{da2018coupling} \\
\hline
\end{tabular}
\end{center}
\end{table*}
This points toward a possibly dominant role of charge order in describing the physics of high critical temperature superconductors. On the one hand, these fluctuations appear indeed  {\itshape universal} among {\itshape all} the cuprate families, including the ``214'' La-based one characterized by spin-charge stripes. On the other hand, they are {\itshape dynamical}, which might help to explain either the superconducting or the strange metal regions, both pervaded by their presence. In this respect, the energies of these charge modulations, in the meV range, are comparable not only  to the superconducting energy gap, but also to the lattice excitations that are probably playing a role, although non exclusive, in Cooper pairing. \cite{dal2012disentangling, johnsoton2010EPCandSC, scalapino2012spinSCpairing}. 

In some cases, charge density features have been observed also at energies of hundreds of meV, hinting at a possible coupling between electronic and magnetic excitations and have been discussed as a possible signature of the pseudogap \cite{yu2020unusual}. However, it is worth highlighting that the high temperature charge fluctuations are generally observed above  $T^*$. This is a strong argument against the picture of the charge order as a secondary effect generated by the pseudogap. On the contrary, it is possible that short-ranged charge modulations are central to the pseudogap phenomenon, possibly being involved in its formation at high temperatures.

These novel results are indeed remarkable advances with respect to the grand challenges of HTS but they also call for further  developments in theory and experiments.
From the theoretical point of view, the general Landau models that, for a Fermi liquid, predicted CDW in the proximity of some form of charge instability ending at zero temperature into a quantum critical point, and that can explain some of anomalous behaviors of cuprates, need to be refined and partly reformulated \cite{castellani1995singular, castellani1996non}. In particular, the lattice and spin degrees of freedom must be explicitly considered in order to get a more realistic picture. Only by taking into account the intertwining of electronic excitations with phonons and magnons one might be able to capture all the consequences of charge order in HTS, including the possible, fundamental, link to the superconducting pairing \cite{van2009optical, muschler2010electron}. 

On the experimental side RIXS has emerged as a superior spectroscopic technique for the  investigation of the multiple dimensions (energy, periodicity, correlation length) and characters (mainly charge, but also lattice and spin) of charge fluctuations.  Non resonant diffraction is blind to the faint charge density modulations and can detect them only when sizable atomic displacements are present, i.e., only for strong CDW. { And EI-RXS, despite a simpler experimental apparatus and higher count rate, suffers from being dazzled by an overwhelming inelastic signal: it can still be viewed as a powerful complementary technique, but it must follow a preliminary RIXS investigation to avoid ambiguous results and risk of misinterpretation of the data.} High resolution RIXS not only allows a proper elimination of higher energy loss excitations (paramagnons, $dd$, charge transfer) but also provides a detailed momentum/energy map of charge fluctuations and phonons. The recent RIXS results presented in this article are at the state of the art \cite{brookes2018beamline}, with energy and momentum resolution \emph{circa} 30-50 meV and 0.01 \AA$^{-1}$ (0.005 r.l.u.) respectively at the Cu $L_3$ edge (930 eV). However, the full comprehension of the charge order phenomenon, and of its connection to the other degrees of freedom, which are responsible for the HTS ground state, asks for further progress in resolution, signal intensity, selectivity and sample environment. Narrower instrumental bandwidth would allow resolving charge density fluctuations from phonons and disentangle effects related to the opening of the superconducting gap \cite{suzuki2018probing},  eventually providing a direct firm determination of the charge fluctuation energy. A more extensive use of the polarization analysis in the RIXS spectra, which at present is only possible at the ID32 beamline of ESRF \cite{brookes2018beamline}, would provide the evidence of the possible entwining of charge and spin fluctuations beyond the electron doped case of Ref. \citenum{da2018coupling}. Lower temperatures for the sample would allow to measure RIXS in the superconducting state also for underdoped and overdoped compounds with $T_{\mathrm{c}}$ lower than 20\,K. And magnetic fields of several tesla would offer the opportunity to study the effects of a weakened superconducting state on charge fluctuations. New opportunities are also offered by RIXS measurements at X-ray free electron lasers (XFEL in Hamburg, LCLS II in Stanford). By pumping with optical pulses of appropriate wavelength \cite{fausti2011light} charge fluctuations and relevant phonons would be excited and RIXS could probe the charge order in the metastable state, and also detect possible simultaneous effects on spin and orbital excitations. On the longer perspective, innovative use of the coherence of the XFEL beams can be envisaged for probing the temporal evolution of charge fluctuations, although coherent scattering experiments seem for the moment unfeasible as they require signals much stronger than that of charge order at high temperature \cite{chen2016remarkable,shen2021snapshot}.  

\begin{acknowledgments}
We thank Lucio Braicovich, Sergio Caprara, Carlo Di Castro, Marco Grilli and Marco Moretti Sala for the many enlightening and lively discussions on this subject. R.A. acknowledges financial support from the Swedish  Research Council  (VR), under the project 2020-04945. G.G. acknowledges the support by project PRIN2017 ``Quantum-2D'' ID 2017Z8TS5B of the Ministry for University and Research (MIUR) of Italy.
\end{acknowledgments}

\bibliography{biblio}

\end{document}